\documentclass{article}

\usepackage{graphicx} 

\usepackage{amsmath,amssymb}
\usepackage{bezier,natbib}
\usepackage{etex}
\usepackage{color}
\usepackage[latin1]{inputenc}
\usepackage{enumerate}

\def\IR{\mathbb{R}}

\def\bmx{{\mathbf x}}

\newtheorem{definition}{Definition}
\newtheorem{proposition}{Proposition}

\newtheorem{theorem}{Theorem}

\renewcommand{\phi}{\varphi}

\bibpunct{(}{)}{;}{a}{,}{,}

\pdfoutput=1

\begin{document}
\title{General notions of depth for functional data}
\author{Karl Mosler \and Yulia Polyakova}
\date{June 2012\\[1mm] revised January 2018}

\maketitle

\begin{abstract}
A data depth measures the centrality of a point with respect to an empirical distribution.
Postulates are formulated, which a depth for functional data should satisfy, and a general approach is proposed to construct multivariate data depths in Banach spaces. The new approach, mentioned as $\Phi$-depth, is based on depth infima over a proper set $\Phi$ of $\IR^d$-valued linear functions. Several desirable properties
are established for the $\Phi$-depth and a generalized version of it.
The general notions include 
many new ones as special cases.
In particular a location-slope depth and a principal component depth are introduced.

\end{abstract}

{\it Keywords:}\/ {Multivariate functional depth, infimum depth, central regions, trimmed regions, $\Phi$-depth, graph depth, location-slope depth, grid depth, principal component depth}.

{\it Address of the authors:\/}
Statistics and Econometrics, Universit{\"a}t zu K{\"o}ln, Albertus Magnus Platz, 50923 K{\"o}ln, Germany,\\
(E-mail: {\tt mosler@statistik.uni-koeln.de})

\section{Introduction}\label{sec1}

In multivariate data analysis a depth function measures how `deep' a point is located in a given data cloud in Euclidean $d$-space, that is, how close it is to an implicitly defined `center' which, in turn, has maximal depth.
Notions of data depth in $\IR^d$ are
closely related to those of multivariate quantiles, central ranks and outlyingness.
The upper level sets of a depth function form central regions that reflect the location, scale and shape of the given distribution.
By this, data depth has become a powerful tool of nonparametric analysis in $\IR^d$.

Many notions of multivariate data depth have been proposed in the literature, starting with \cite{Tukey75} and \cite{Liu90}. They have been successfully applied to problems of
nonparametric statistical analysis in $\mathbb{R}^d$ such as classification (or supervised learning), hypothesis testing, and others.
A theory of depth functions in $\IR^d$ has been developed that includes population versions (that is, depth with respect to a probability distribution) as well as basic postulates and other properties shared by these notions; see \cite{ZuoS00a}, \cite{Mosler02a}, \cite{Dyckerhoff04} and the surveys by \cite{Serfling06} and \cite{Cascos09}. For exemplary applications refer to
\cite{LiuPS99}, \cite{LiCAL12}, and \cite{DuttaG12}.
Also several depth notions have been proposed for functional data, e.g., by \cite{FraimanM01}, \cite{CuestaAN08b}, \cite{LopezPR09}, and \cite{ClaeskensHSV14}. Intended applications include problems of classification, outlier detection and the trimming of functional data.
However, extending a multivariate data depth to a Banach space $E$ of infinite dimension causes substantial problems.
\cite{DuttaGC11} show that in standard settings the Tukey functional data depth collapses to zero with probability one.
The reason is that the dual space $E'$ is too large, or, put another way, the unit ball
of $E$ is not compact.

What is still missing is a theory of depth functions in functional spaces, in particular a general definition based on proper postulates to be imposed on such a depth.
The present paper contributes to this issue in two respects. First, by formulating a set of minimal postulates for a depth function in a Banach space $E$ and, second, by introducing a comprehensive class of functional infimum depths, named $\Phi$-depths. Here $\Phi$ consists of linear functions that map $E$ to a finite-dimensional space. Each $\phi\in \Phi$ may be regarded as a particular `view' on the functional data or `aspect' of them, and the depth is defined as the infimum of depths regarding these (multivariate) views. An aspect of a function can, e.g., be its projection to some finite-dimensional marginal, particularly its values at one or several times, a derivative, or a mean value of the function in a subinterval.

The postulates given below are weak enough to generate nontrivial depth notions. They are contrasted with further postulates, weaker as well as stronger ones.
Especially, in the case of symmetrically distributed data, the postulates imply that the depth takes its maximum at the center of symmetry.
A functional data depth $D$ generates central regions $D_\alpha$, consisting of all functions that have at least a certain depth $\alpha\in[0,1]$. These regions  describe the data cloud regarding its location, variation and functional shape; they satisfy similar postulates.

By specializing the set $\Phi$ many different notions of depth are obtained. As important subclasses of $\Phi$-depths we introduce the general graph depths and the grid depths. 
The location-slope graph (or grid) depth is a bivariate depth that operates on functions and their first derivatives simultaneously. It can be used to analyze warped functional data by incorporating their slope together with a warping function that has been estimated from the data.
Two extensions of the $\Phi$-depth suggest themselves: to take a \emph{weighted} infimum of $d$-variate depths and to make the set $\Phi$ dependent on the data. Both extensions come out to be compatible with the basic postulates. The latter gives rise to the notion of principal component depth, where
an $m$-variate data depth is applied to the loadings of the first $m$ principal components.

Overview of the paper: Section \ref{sec2} gives a short account of data depth in finite dimensions and its basic properties. In Section \ref{sec3} a set of postulates is formulated that define a general functional data depth; then the class of $\Phi$-depths is introduced and its properties are derived. Next, in Section \ref{sec3a}, additional postulates are given that may be satisfied in special cases. Section \ref{sec4} discusses restrictions to be imposed on $\Phi$ that are specific to the functional data setting. Special classes of $\Phi$-depths are discussed in Section \ref{sec5}. Section \ref{sec6} presents the generalized $\Phi$-depth and, in particular, the principal component depth, while Section \ref{sec6a} gives an outlook on population versions. Section \ref{sec7} concludes with alternative approaches.

\section{Multivariate data depth}\label{sec2}
First let us recapitulate the notion of a depth for data in finite-dimensional space $\mathbb{R}^d$.
A \textit{multivariate ($d$-variate) data depth} is a bounded function that, to a given data cloud $X=\{x^1, \dots, x^n\}\subset \mathbb{R}^d$ and a point $y\in \mathbb{R}^d$, $d\ge 1$, assigns a depth value $D(y|X)=D(y |x^1, \dots, x^n)\in \mathbb{R}$ that satisfies certain postulates, that is, desirable properties.  For the present inquiry we use the following set of postulates, which is due to  \cite{Dyckerhoff02a}:
\begin{itemize}
  \item \textbf{D1 Translation invariant:} $D(z+b|x^1+b, \dots, x^n+b)=D(z|x^1, \dots, x^n)$ for all $b\in \mathbb{R}^d$\,,
  \item \textbf{D2 Linear invariant:} $D(Az|Ax^1, \dots, Ax^n)=D(z|x^1, \dots, x^n)$ for every regular matrix $A\in \mathbb{R}^{d\times d}$\,,
  \item \textbf{D3 Null at infinity:} $\lim_{\left\|z\right\|\rightarrow\infty}D(z|x^1, \dots, x^n)=0$\,.
  \item \textbf{D4 Monotone on rays:} If a point $z^*$ has maximal depth, that is $D(z^*|x^1, \dots, x^n)=\max_{z\in\mathbb{R}^d}D(z|x^1, \dots, x^n)$\,,
  then for any $r$ in the unit sphere ${S}^{d-1}$ the function $\beta\mapsto D(z^*+\beta r|x^1, \dots, x^n)$ decreases with $\beta>0$\,,
  \item \textbf{D4con Quasiconcave:} $D(\cdot|x^1, \dots, x^n)$ is a quasiconcave function, that is, its upper level sets $D_\alpha(x^1,\dots, x^n) =  \{z\in\mathbb{R}^d : D(z|x^1,\dots, x^n) \geq \alpha\}$  are convex for all $\alpha >0$\,.
  \item \textbf{D5 Upper semicontinuous:} The upper level sets  $D_\alpha(x^1,\dots, x^n)$ are closed for all $\alpha >0$\,.
\end{itemize}

Slightly different postulates have been given by \cite{Liu90} and  \cite{ZuoS00a}. The main difference between these postulates and those above is that they   refer to a center of symmetry at which depth should attain its maximum and that they do not require upper semicontinuity (which serves as a useful technical restriction).
Clearly, {\bf D4} implies that if $X$ is centrally symmetric then $D(\cdot|X)$ attains its maximum at the center of symmetry.
At the end of Section \ref{sec3a} we will come back to the behavior of a functional depth under symmetry.

For $\alpha\ge 0$ the level sets $D_\alpha(x^1,\dots, x^n)$ form a nested family. They are mentioned as \textit{depth trimmed regions} or \textit{central regions}, with $\alpha$ measuring the degree of centrality. The above postulates can be equivalently formulated in terms of these regions. \textbf{D1} and \textbf{D2} say that the family of central regions is equivariant against shifts and changes of scale, respectively. \textbf{D3} means that for any $\alpha>0$ the region $D_\alpha(x^1,\dots, x^n)$ is bounded. \textbf{D4} states the starshapedness of each $D_\alpha(x^1,\dots, x^n)$ with respect to $z^*$.
\textbf{D4con} and \textbf{D5} say that each region is convex and closed, respectively. Obviously, as a convex set is starshaped with respect to each of its points, \textbf{D4con} implies \textbf{D4}.

Depth trimmed central regions describe a data cloud $x^1, \dots, x^n\in \IR^d$ with respect to location, dispersion, and shape. This has many applications in multivariate data analysis as well as inference; see, e.g., \cite{LiuPS99} and the survey by \cite{Serfling06}.
By definition a $d$-variate data depth is bounded. If there is a point of maximum depth, this depth will w.l.o.g.\ be set to 1. Then the innermost level set arises at $\alpha=1$, and $D_1(x^1,\dots,x^n)$ is the set of \textit{deepest points}.

More general, in place of the data cloud a probability measure $\mu$ or a random vector $X$ on $\mathbb{R}^d$ can be considered.
This is mentioned as a \textit{multivariate depth}. In turn, applied to an empirical distribution $\mu$ that gives equal probabilities to the points $x^1, \dots, x^n$, a multivariate data depth is obtained.

Important examples of multivariate depth functions are, among many others, the Mahalanobis depth, the Tukey (or halfspace) depth,
the simplicial depth
\citep{Liu90},
the projection depth (\cite{Liu92}, \cite{ZuoS00a}),
and the zonoid depth \citep{KoshevoyM97b}.
The various depths proposed in the literature all satisfy the postulates \textbf{D1} to \textbf{D3} (some only orthogonal invariance in place of linear one), while the remaining postulates are met to a different extent.
They show different additional features regarding their practical applicability (like computability and robustness) as well as analytical properties (like continuity and characterization through marginals), which permit their application in different statistical tasks. In particular, efficient algorithms are needed to perform bootstrap tests (\cite{Dyckerhoff02}) or high-dimensional classification tasks based on data depth (\cite{LangeMM14a,LangeMM14}).

\section{Functional data depth: Postulates, $\Phi$-depth}\label{sec3}

Consider a Banach space $E$ having some norm $||\cdot||$. Let $E'$ be the dual space of all continuous linear functionals endowed with the operator norm $||\phi||'=\sup_{||x||=1}|\phi(x)|$, $B_E$ and $B_{E'}$ be the unit balls ($S_E$ and $S_{E'}$ the unit spheres) of $E$ and $E'$, respectively, and ${\cal B}$ the Borel sets of $E$.
Prominent and practically relevant examples for $E$ include spaces of functions\footnote{For an $\IR^d$-valued function $x$ we notate $x(t)=(x_1(t),\ldots,$ $x_d(t))$.} mapping a compact interval $J$ to Euclidean space $\IR^d$, in particular:
\begin{itemize}
  \item the space $C(J)$ of real-valued continuous functions $x:J\to \mathbb{R}$ with a norm $||x||_{\infty}=\sup_{t\in J}|x(t)|$,
  \item the space $L^2(J)$ of real-valued square-integrable functions $x:J\to \mathbb{R}$ with a norm $||x||_2=(\int_J x^2(t) dt)^{1/2}$,
  \item the space of $\mathbb{R}^d$-valued continuous functions on $J$,  $C\left(J; \mathbb{R}^d\right)$, endowed with the norm $||x||_{\infty}=\sup_{t\in J}||x(t)||$, where $||\cdot||$ is an arbitrary norm on $\mathbb{R}^d$,
  \item the space of real-valued $m$-times continuously differentiable functions on $J$, $C^m(J)$, with the norm
       $||x||^{(m)}=max_{s\le m}\{|| x^{(s)}||_\infty\}$.
  \item the space ${\ell}^p$ of sequences $x=(x(t))_{t\in \mathbb{N}}$ in $\mathbb{R}$ that have finite norm $||x||_{p}=\left(\sum_{t=1}^\infty |x(t)|^p\right)^{\frac1p}$, for some $p$, $1\le p < \infty$,
\end{itemize}


A \emph{functional data depth} is a real-valued functional that, given a finite data cloud of elements in $E$, indicates how `deep' another given element of $E$ is located in the data cloud, that means, how `close' it is to the `center' of the cloud. Of course, the meaning of `deep', `close' and `center' are implicitly determined by the functional depth.

 We formulate general postulates which a meaningful definition of functional depth should reasonably satisfy and check their eventual restrictions implied by them on the class $\Phi$ in Definition \ref{defFunctionalDepth}. For short we will notate the data clouds by  $X=\{x^1, \dots, x^n\}$, $\phi(X)=\{\phi(x^1), \dots, \phi(x^n)\}$, and similarly
$AX= \{A x^1,\dots A x^n\}$, $\lambda X= \{\lambda x^1,\dots \lambda x^n\}$, etc..

 Our postulates extend the multivariate postulates \textbf{D1} to \textbf{D5} to the general setting; they involve elements of an arbitrary Banach space.  Further postulates are given below that are specific to spaces of functions on a bounded real interval.

\begin{itemize}
  \item \textbf{FD1 Translation invariant:} $D(z+b|X+b)=D(z|X)$ for all $b\in E$\,.
  \item \textbf{FD2 Scale invariant:} $D(\lambda z|\lambda X)=D(z|X)$ for all $\lambda > 0$\,.
  \item \textbf{FD3 Null at infinity}\textbf{:} $\lim_{||z|| \rightarrow\infty, z\in S}D(z|X)=0$\,, where $S$ is a certain fixed subspace $S$ of $E$.
  \item \textbf{FD4 Monotone on rays:} For any
  $z^*$ with $D(z^*|X)=1$
  and any $r\in B_E$ the function $\beta\mapsto D(z^*+\beta r|X)$ decreases with $\beta>0$\,.
   \item \textbf{FD4con Quasiconcave:} The upper level sets $D_\alpha(X) =
  \{z\in E : D(z|X) \geq \alpha\}$  are convex for all $\alpha >0$.
  \item \textbf{FD5 Upper semicontinuous:} The sets $D_\alpha(X)$ are closed for all $\alpha >0$\,.
\end{itemize}
The invariance postulates \textbf{FD1} and \textbf{FD2} say, which aspects of the data a functional depth should \emph{not} reflect.
\textbf{FD3} postulates that the depth of a function should vanish if, in a certain subspace $S$ of $E$, its norm goes to infinity; in other words, the intersection of an upper level set (= trimmed region) of the depth with $S$ should be bounded. As the postulate depends on $S$ we may also explicitly write \textbf{FD3}$(\mathbf S)$.
\textbf{FD4} essentially says that the depth function is unimodal and has its maximum at some central point $z^*$.
Again, \textbf{FD4con} is stronger than \textbf{FD4}. A functional depth that satisfies the stronger postulate \textbf{FD4con} is named a \emph{convex depth}. \textbf{FD5}, like \textbf{D5}, is a technical assumption.

The given postulates correspond to properties of the trimmed regions that originate from a functional depth.
  We provide a list of postulates on the family $\{D_\alpha(X)\}_\alpha$ which are equivalent to the above postulates \textbf{FD1} to \textbf{FD5} on the depth $D$: For all $X=\{x^1,\dots,x^n\}\in E$ and $\alpha\ge 0$,
\begin{itemize}
            \item \textbf{FR1:} $D_\alpha(X+b)=D_\alpha(X)+b\,, \quad b\in E\,,$
            \item \textbf{FR2:} $D_\alpha(\lambda X)=\lambda D_\alpha(X)\,, \quad \lambda >0\,,$
            \item \textbf{FR3:} $D_\alpha(X)\cap S$ is bounded for all $\alpha>0$, where $S$ is a certain fixed subspace $S$ of $E$\,,
            \item \textbf{FR4:} $D_\alpha(X)$ is starshaped with respect to all $z^*\in \bigcap_{\phi \in \Phi} \phi^{-1}(D_1^d(\phi(X)))$\,,
            \item \textbf{FR4C:} $D_\alpha(X)$ is convex,
            \item \textbf{FR5:} $D_\alpha(X)$ is closed.
\end{itemize}

Further, for any family of depth trimmed central regions we obtain the equation
\begin{equation}\label{intersectionprop}
D_\alpha(X)= \bigcap_{\beta<\alpha} D_\beta(X)\,, \quad \alpha\in ]0,1]\,,
\end{equation}
which is obvious as it holds for the upper level sets of any function.

As a first simple example  of a functional data depth in the sense of postulates \textbf{FDi} or, equivalently, \textbf{FRi} for $I=1,2,3,4,5$, we consider the \emph{half-region depth} of \cite{LopezPR11}. The half-region depth (previously introduced as \emph{half-graph depth} in \cite{LopezPR05}) of a function $z$ with respect to a data cloud $X=\{x^1, \dots, x^n\}$ is defined as
\begin{equation}\label{HRdepth}
  D^{halfregion}(z|X) = \min \{P_n(X\le z), P_n(z \le X)\},
\end{equation}
where $P_n$ is the empirical measure on $X$ and $\le$ denotes pointwise ordering of functions.
It is immediately seen that postulates \textbf{FD1} and \textbf{FD2} are satisfied. Also, when $E$ is the space of real functions defined on $[0,1]$ and equipped with the supremum norm, \textbf{FD3} is met for $S=E$ . Observe that the $\alpha$-central region is
\[ D^{halfregion}_\alpha=\left\{z : \frac 1n \sum_{i=1}^n I(x^i\le z)\right\} \cap \left\{z : \frac 1n \sum_{i=1}^n I(z \le x^i)\right\},
\]
which is a closed set, hence \textbf{FR5} holds and, equivalently, \textbf{FD5}. Finally, to demonstrate \textbf{FR4}, note that the median set $A^*= argmax_z D^{halfregion}(z|X)$ is empty. Let $z^*\in A^*$, and w.l.o.g. (due to \textbf{FR1}) assume $z^*=0$. Then, in considering $D^{halfregion}(\beta r|X)$ distinguish three cases: $r\le 0$, $0\le r$, and $r$ has positive and negative values. In each of these cases it is obvious that $D^{halfregion}(\beta r|X)$ decreases with $\beta$ as less of the $x^i$ are completely below resp. above $\beta r$; hence \textbf{FR4} is satisfied.

In the sequel we investigate functional depths of a general infimum form that is given in Definition \ref{defFunctionalDepth}. This definition includes several notions of functional data depth that are known from the literature as well as many new ones.
Some other existing notions, that are not covered by our definition, will be addressed  in Section \ref{sec7} below.

\begin{definition}[$\Phi$-depth]\label{defFunctionalDepth}
For $z, x^1,\dots, x^n\in E$, define
\begin{equation}\label{defFD}
 D(z|x^1, \dots, x^n)= \inf_{\phi\in \Phi} D^d(\phi(z)|\phi(x^1), \dots, \phi(x^n))\,,
\end{equation}
where $D^d$ is a $d$-variate data depth satisfying {\rm \textbf{D1}}, {\rm \textbf{D2}}, {\rm \textbf{D3}}, {\rm \textbf{D4}}, and {\rm \textbf{D5}},  $\Phi\subset {E'}^d$, and ${E'}^d$ is the space of continuous linear functions $E\to \IR^d$. $D$ is called a \emph{$\Phi$-depth}. More shortly, we write
\begin{equation*}
D(z|X)= \inf_{\phi\in \Phi} D^d(\phi(z)|\phi(X))\,.
\end{equation*}
\end{definition}

Each $\phi$ in our definition may be regarded as a particular aspect we are interested in and which is represented in $d$-dimensional space. The
depth of $z$ is given as the smallest multivariate depth of $z$ under all these aspects. It implies that all aspects are equally relevant so that the depth of $z$ cannot be larger than its depth under any aspect. For example, $\phi(z)$ may be the evaluation of a function $z\in L^2(0,1)$ at one or several arguments $t\in[0,1]$.

As the $d$-variate depth $D^d$ takes its maximum at 1, the functional data depth $D$ is bounded above by 1. At every point $z^*$ of \textit{maximal $D$-depth} it holds $D(z^*|X)\le 1$. The bound is attained with equality, $D(z^*|X)= 1$, iff
$D^d(\phi(z^*)|\phi(X))=1$ holds for all $\phi\in \Phi$, that is, iff
\begin{equation}\label{deepest}
z^* \in \bigcap_{\phi \in \Phi} \phi^{-1}(D_1^d(\phi(X)))\,.
\end{equation}

Now we proceed to our first main result, which says that a $\Phi$-depth indeed satisfies the relevant postulates.

\begin{theorem}\label{theo1}
\begin{description}
  \item[(i)] A $\Phi$-depth (\ref{defFunctionalDepth}) always satisfies {\rm \textbf {FD1}}, {\rm \textbf {FD2}}, {\rm \textbf {FD4}}, and {\rm \textbf {FD5}}.
  \item[(ii)] The depth  satisfies {\rm \textbf {FD3}} if for every sequence $(z^i)\subset S$ with $||z^i||\to \infty$  exists a sequence $(\phi_i)$ in $\Phi$ such that $||\phi_i(z^i)||\to \infty$\,.
\item[(iii)] It satisfies {\rm \textbf{FD4con}} if the underlying $d$-variate depth satisfies {\rm \textbf{D4con}}.
\end{description}

\end{theorem}

The proof is given in the Appendix. Obviously, for \textbf {FD3} to hold, $\Phi$ must be rich enough.
When the functional depth is univariate ($d=1$) and all functions in $\Phi$ are increasing, an additional monotonicity postulate is satisfied, which refers to the pointwise order of functions:

Regarding the central regions of a $\Phi$-depth we have the following result.
\begin{theorem}\label{weaksets}
Let $D$ be a functional depth $D$ according to Definition (\ref{defFD}).
For the level sets $D_\alpha$ of $D$ it holds
\begin{equation}\label{weakpropsets}
D_\alpha(X)= \bigcap_{\phi\in\Phi} \phi^{-1}(D^d_\alpha(\phi(X)))\,.
\end{equation}
\end{theorem}

The theorem is proven in the Appendix. It has consequences for the calculation of central regions of functions: $D_\alpha(X)$ may be approximately determined by computing a sequence of $d$-variate central regions and taking their intersection. Observethat the right hand side in Equation (\ref{weakpropsets}) can be empty for some $\alpha_0\le 1$ and, consequently, for all $\alpha\in [\alpha_0,1]$. This happens also  if $E=\IR^m$ and $d=1$, i.e., if the depth (\ref{defFD}) is a multivariate depth; e.g., the Tukey depth.    $D_\alpha$ is nonempty for all $\alpha\in [0,1]$ iff the set
$\bigcap_{\phi \in \Phi} \phi^{-1}(D_1^d(\phi(X)))$ is not empty,
that is, some $z^*$ exists satisfying (\ref{deepest}).

\section{Further postulates}\label{sec3a}

Postulate \textbf{FD2} may be strengthened to full linear invariance,

\begin{itemize}
\item \textbf{FD2L Linear invariant:} $D(Az|AX)=D(z|X)$ for every isomorphism $A:E\to E$
(that is, for every linear continuous transformation $A:E\to E$ that has a continuous inverse),
\end{itemize}
which corresponds to the postulate \textbf{D2} of multivariate depth.
\textbf{FD2L} appears to be a rather strong restriction; it implies that the depth is invariant against all linear isometries of $E$. Moreover,
when $E$ is a function space
like $C(J)$
with $||\cdot||_\infty$, e.g.\ all transformations of type
$A_K x = x-\int_0^1 K(\cdot,t)x(t) dt$ (with some kernel $|K|\le 1$) are included in the invariance postulate. Also, the depth then is invariant against any rearrangement $A_\rho$ of the functions, $A_\rho x(t)= (x\circ \rho)(t))$, with an arbitrary bijection $\rho:J\to J$. We formulate the latter as a two separate postulates:
\begin{itemize}
\item \textbf{FD2R Rearrangement invariant:} \\ $D(z\circ \rho | X\circ \rho ) =
 D(z|X)$ for every bijective function $ \rho: J\to J$\,,
\end{itemize}
\begin{itemize}
    \item \textbf{FD2IR Increasing rearrangement invariant:} $D(z\circ \rho | X\circ \rho)=D(z|X)$ for every increasing bijective function $ \rho: J\to J$\,.
\end{itemize}
Clearly, \textbf{FD2R} implies \textbf{FD2IR}. The property \textbf{FD2} means that the development of a function in time $t$ is irrelevant to the depth, while the property \textbf{FD2R} indicates that the speed of elapsing time plays no role.

We continue with another postulate that is specific to function spaces. Let $E$ be a Banach space of functions $J\to \IR^d$, where $J$ is a real interval, and notate by $E_0$ the subset of functions $a\in E$ that vanish nowhere $a(t)\not= 0$ for all $t\in J$.
\begin{itemize}
    \item \textbf{FD2F Function-scale invariant:} If $E$ is equipped with a product, say, the pointwise product of functions, it holds $D(a\cdot z|a\cdot X)=D(z|X)$ for all $a \in E_0$\,.
\end{itemize}
Obviously, both function-scale invariance \textbf{FD2F} and full linear invariance \textbf{FD2L} are stronger than scale invariance \textbf{FD2}.


\begin{proposition}
Let $d=1$ and all $\phi\in\Phi$ be increasing.  Then the $\Phi$-depth (\ref{defFunctionalDepth}) satisfies the postulate {\rm \textbf{FD4pw}},
\end{proposition}
\begin{itemize}
  \item \textbf{FD4pw Monotone:} For every $z^*\in E$ having maximal depth it holds that
$D(y|X) \ge D(z|X)$ whenever either $z^*\le y \le z$ or $z\le y \le z^*$\,.
\end{itemize}

\textbf{Proof.}
Let $z^*$ have maximal depth and assume $z^*\le y \le z$. Then $\phi(z^*)\le \phi(y)\le \phi(z)$ for all $\phi$ (as $\phi$ is increasing) and
\[ D^1(\phi(y)|\phi(X)) \ge D^1(\phi(z)|\phi(X))
\]
due to \textbf{D4}. Taking the infimum on the left and right hand sides yields   $D(y|X) \ge D(z|X)$\,. The case
$z^*\ge y \ge z$ is similarly shown.
\hfill $\lozenge$

As it was mentioned above, other sets of postulates for a multivariate depth require that the depth should be maximal at some center of symmetry.
Different notions of symmetry with respect to the origin have been considered in the context of special data depths:  central symmetry, angular symmetry \citep{Liu90}, and halfspace symmetry \citep{ZuoS00a}. Consider the following postulate:
\begin{itemize}
  \item {\bf FD4center:} If  $X$ is symmetrically (in a proper sense) distributed about some $z^c$, then the depth is maximal at $z^c$.
 \end{itemize}

As our notion of functional data depth is based on linear functions $\phi$ into $\IR^d$, the maximality of a $d$-variate data depth at a point of either central or angular (or halfspace) symmetry carries over immediately to the same maximality of the functional data depth; this holds for any choice of $\Phi$:

\begin{proposition}
   If $X$ is centrally symmetric about some $z^c\in E$ and there exists some $ z^*\in E$ with  $D(z^*|X)=1$, then the  $\Phi$-depth $D$ is maximal at $z^c$. The same holds in the cases of angular (halfspace) symmetry if the corresponding $D^d$ attains its maximum at the center of angular (resp.\ halfspace) symmetry.
\end{proposition}


In terms of functional data depth: If the data is symmetric about some function, then this function is deepest. If the data is not symmetric, different deepest functions arise depending on the choice of $D^d$ and $\Phi$; see Section 5 below.

\section{Restrictions on $\Phi$}\label{sec4}

We now proceed with specifying the set $\Phi$ of functions and the multivariate depth $D^d$ in (\ref{defFD}). While many features of our functional data depth (\ref{defFD}) resemble those of a multivariate depth, an important difference must be pointed out:
In a general Banach space the unit ball $B$ is not compact, and properties \textbf{FR3} and  \textbf{FR5} (or equivalently  \textbf{FD3} and  \textbf{FD5}) do not imply that the level sets of a functional data depth are compact.

 We start with an example which demonstrates the need for imposing strong restrictions on $\Phi$.

\textbf{Example (Tukey functional depth):}
Let $E$ be an infinite dimensional Banach space. With $\Phi$ being the whole dual space, $\Phi=E'$, and $D^1$ the univariate Tukey depth $TD^1$, for $\zeta, \xi_1,\dots, \xi_n\in \mathbb{R}$,
\begin{equation}\label{defHalfspaceFunctional}
TD^1(\zeta|\xi_1,\dots, \xi_n)= \frac 1n \min\{\# \{i : \xi_i\le\zeta\}, \# \{i : \xi_i\ge\zeta\}\}\,,
\end{equation}
we obtain the \textit{Tukey functional depth} $TD$. It satisfies all postulates \textbf{FD1} to \textbf{FD5} and even the stronger \textbf{FD4R} and \textbf{FD2L} (as $\phi\in\Phi\ \Rightarrow A\circ \phi \in \Phi$).
However this depth is not really meaningful. First, note that the depth vanishes outside the convex hull of the data, which has affine dimension $\le n-1$. Moreover, consider a measure $\mu$ on $(E,{\cal B})$ that has zero probability on all finite dimensional linear manifolds (e.g. $E=\ell^p$ and $\mu$ an infinite product of $L$-continuous measures on the reals); then $TD(\cdots|\xi_1,\dots, \xi_n)=0$ $\mu$-almost surely in $E$.
That is, the data version of the Tukey functional depth collapses to zero with probability one. \cite{DuttaGC11} provide an example where this happens as well with the population version. See also Remark 2.8 in \cite{KuelbsZ15}.

The reason why the definition of the example collapses is that the dual space $E'$ is too large, or, put another way, the unit ball $B_E$
of $E$ is not compact. So, to obtain a meaningful notion of functional data depth of type (\ref{defFD}) one has to carefully choose a set of functions $\Phi$ that is not too large. On the other hand, $\Phi$ should not be too small, in order to extract sufficient information from the data.

 Each $\phi\in \Phi$ corresponds to an aspect of the data that is transformed into $d$-space. An interesting question is whether the transformation of a depth-trimmed region always coincides with the respective region of the transformed data. For this, we introduce the following restriction, which we call the surjection property.

\begin{definition}[Surjection property]
A functional data depth (\ref{defFD}) has the \textbf{surjection property} if, given a cloud $X$, for every $\phi\in \Phi$ and $y\in \IR^d$ there exists some $z\in \phi^{-1}(y)$ so that
\[ D(z|X)= D^d(y|\phi(X))\,.
\]
\end{definition}

Note that in the special case of $D$ being a $k$-variate data depth ($E=\IR^k$) and $\Phi$ a set of real-valued functionals ($d=1$),
the surjection property becomes the \textit{strong projection property} of depth $D$; see \cite{Dyckerhoff04}.

\begin{theorem}\label{theosurjection}
A {functional data depth} (\ref{defFD}) satisfies the \textbf{surjection property}
if and only if for every $\phi\in \Phi$
\begin{equation}\label{strictpropsets}
\phi(D_\alpha(X))= D^d_\alpha(\phi(X))\,
\end{equation}
\end{theorem}

For proof see the Appendix.
According to this Theorem a central set remains central under every aspect. A point (= function) $z$ is more central than another point $w$ if and only if it is more central under every aspect $\phi$.
If a functional depth $D$ has the surjection property the support function of its central regions can be derived
similar to Theorem 3 in \cite{Dyckerhoff04}.
Examples of functional depths that satisfy the surjection property will be given below; see Section \ref{subsecgraph} and \ref{subsecgrid}.


\section{Special classes of $\Phi$-depths}\label{sec5}

This section presents examples of $\Phi$-depths, where the set $\Phi$ of aspects is chosen in a special way.
The Banach space be a space of functions that live on a bounded interval $J$. In the first group of examples, called graph depths, the aspects correspond to the points of some subset of $J$, which may be finite or not, and the depth of a function is evaluated at all these points. In another group of examples, mentioned as grid depths, a function is evaluated on a $k$-point grid and the aspects correspond to directions in $\IR^k$.
In both approaches derivatives of the function may be included, which results in depths that measure similarities regarding the level as well as the slope and possibly higher derivatives of the function.

\subsection{Graph depths}\label{subsecgraph}

Consider $E=C\left(J;\mathbb{R}^d\right)$ with norm $||\cdot||_\infty$
and let
\begin{equation}\label{bandPhi}
\Phi= \{\phi^t:E\to \IR^d | \phi^t (x)=(x_1(t)\ldots ,x_d(t)), t\in T\}
\end{equation}

for some $T\subset J$, which e.g. may be a subinterval or a finite set in $J$. For $D^d$ use any multivariate depth that satisfies \textbf{D1} to \textbf{D5}. This results in the $d$-variate \emph{graph depth}
\begin{equation}\label{defBand}
 D^G(z|x^1,\dots, x^n)= \inf_{t\in T} D^d(z(t)|x^1(t),\dots, x^n(t))\,.
\end{equation}

\begin{proposition}\label{propositiongraph}
A graph depth satisfies the postulates {\rm \textbf{FDi}}, $i\in\{1,2,4,5\}$, {\rm \textbf{FD3}} with $S=(\mathbb{R}^d)^T$, and, in addition, 
{\rm \textbf{FD2IR}} and {\rm \textbf{FD2R}}.
It satisfies {\rm \textbf{FD4con}} if $D^d$ satisfies  {\rm \textbf{FD4con}}. \end{proposition}

The Proposition \ref{propositiongraph} is proven in the Appendix.

The property \textbf{FD2R} means that the development of functions in time is irrelevant to a graph depth.

In particular, let $T=J$ and $d=1$. 
 For $D^1$ we may employ, e.g., the (univariate) Tukey depth; the resulting graph depth, mentioned as the \emph{Tukey graph depth} and notated as $D^{TG}$, satisfies all five basic postulates.

We illustrate the notion of Tukey graph depth by applying it to a wellknown data set from \cite{RamsayS05}.
Figure 1 
exhibits the data, which describe two angles (hip and knee) of a gait cycle measured over 20 time periods at 39 subjects, and the borders of central regions $D^{TG}_\alpha$ (fat lines)
for $\alpha=.25$ and $\alpha=.5$\,, the latter consisting of a single `deepest' function. The bivariate Tukey depth and the central regions have been computed with the algorithms given in \cite{RousseeuwR96} und \cite{RutsR96}. Figure 2 
presents central regions of the hip data for $\alpha\in\{0.02, 0.08, 0.14, 0.2\}$.
Note that the largest of these regions contains \textit{all} data as $\alpha=0.02<1/39$. Finally, Figure 3 
exhibits those data that are \emph{not} included in the depth trimmed regions, and therefore can be regarded as outliers at different levels $\alpha$.

\begin{figure}
\begin{center}
\includegraphics[scale=0.71]{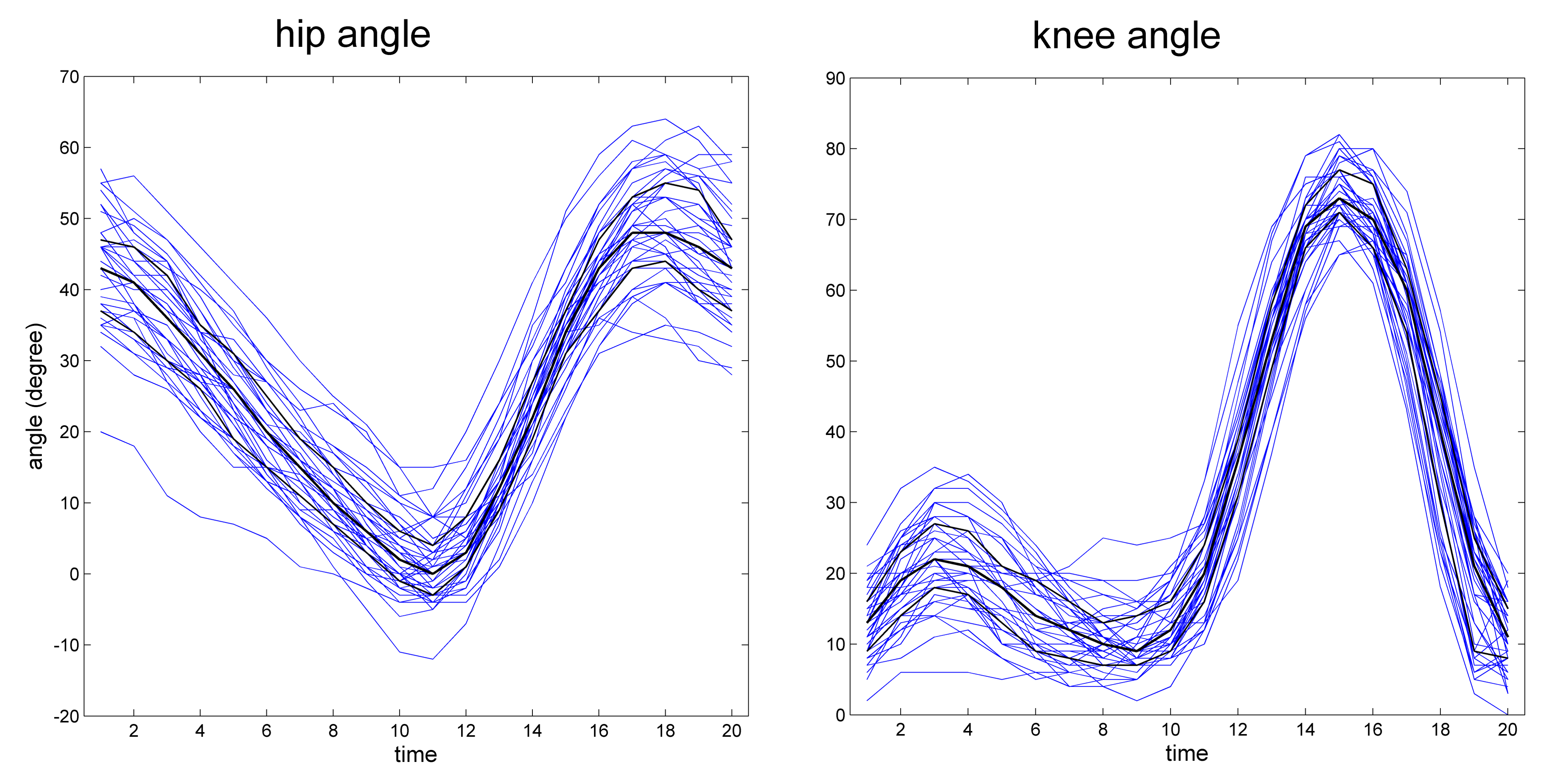}
\caption{Hip resp.\ knee data, with central regions $D^{TG}_\alpha$, $\alpha\in \{0.25,0.5\}$.}
\end{center}
\label{fig1}
\end{figure}

\begin{figure}[h!]
\begin{center}
\label{HZBL}
\includegraphics[scale=0.71]{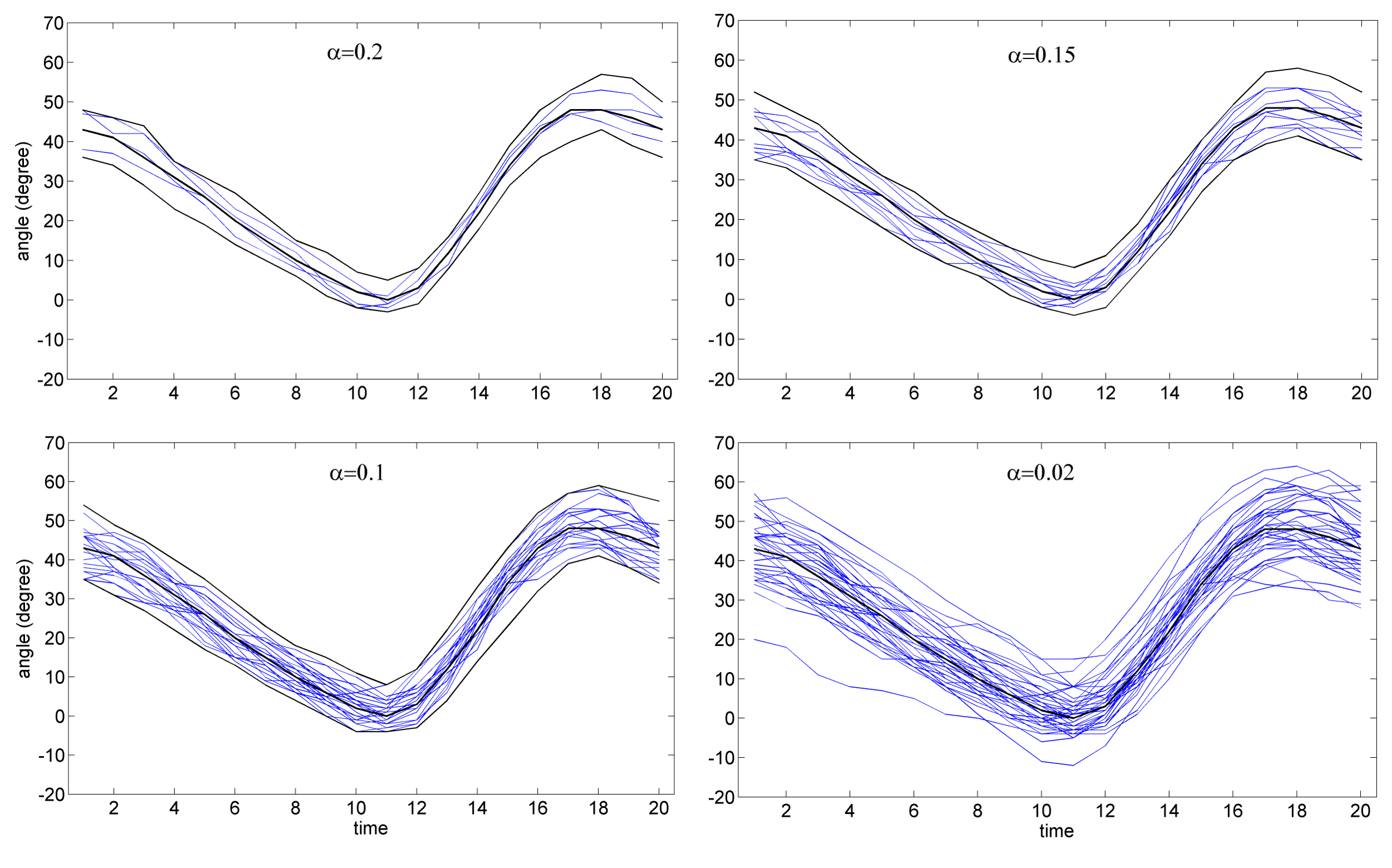}
\caption{Central regions  $D^{TG}_\alpha$, $\alpha\in \{0.02, 0.10, 0.15, 0.20\}$; hip angle.
}
\label{fig2}
\end{center}
\end{figure}

Further, we analyze the above data set in two dimensions: at every time point we have two real values, the hip and the knee angle. We choose the bivariate Tukey depth as the underlying depth $D^2$ in (\ref{bandPhi}). Figure 4 
presents central regions of the resulting bivariate Tukey graph depth for the hip knee data set.

\begin{figure}
\begin{center}
\includegraphics[scale=0.57]{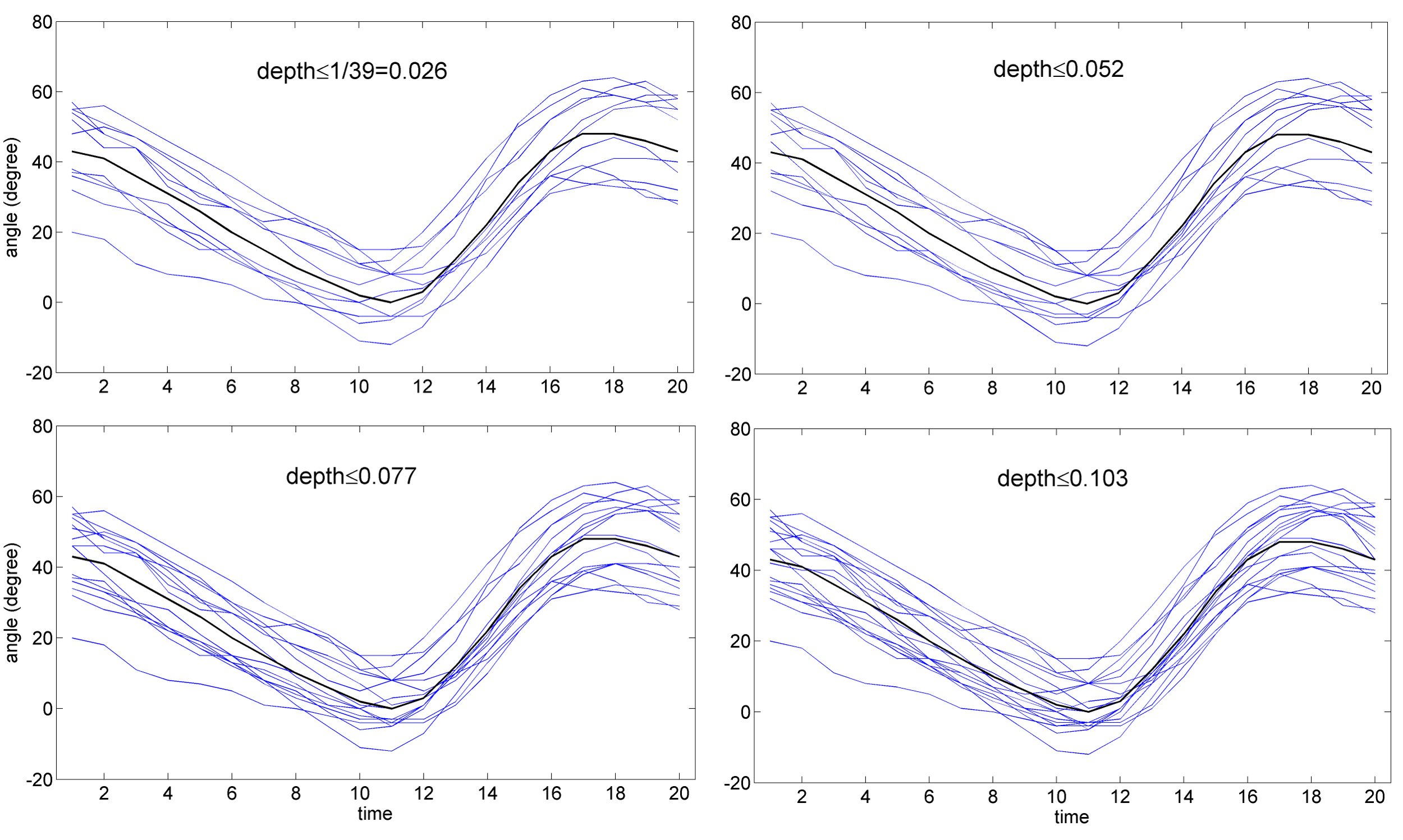}\\
\caption{Outlying data w.r.t.\ $D^{TG}$; hip angle.}
\end{center}
\label{fig3}
\end{figure}

\begin{figure}[h!]
\begin{center}
\includegraphics[scale=0.71]{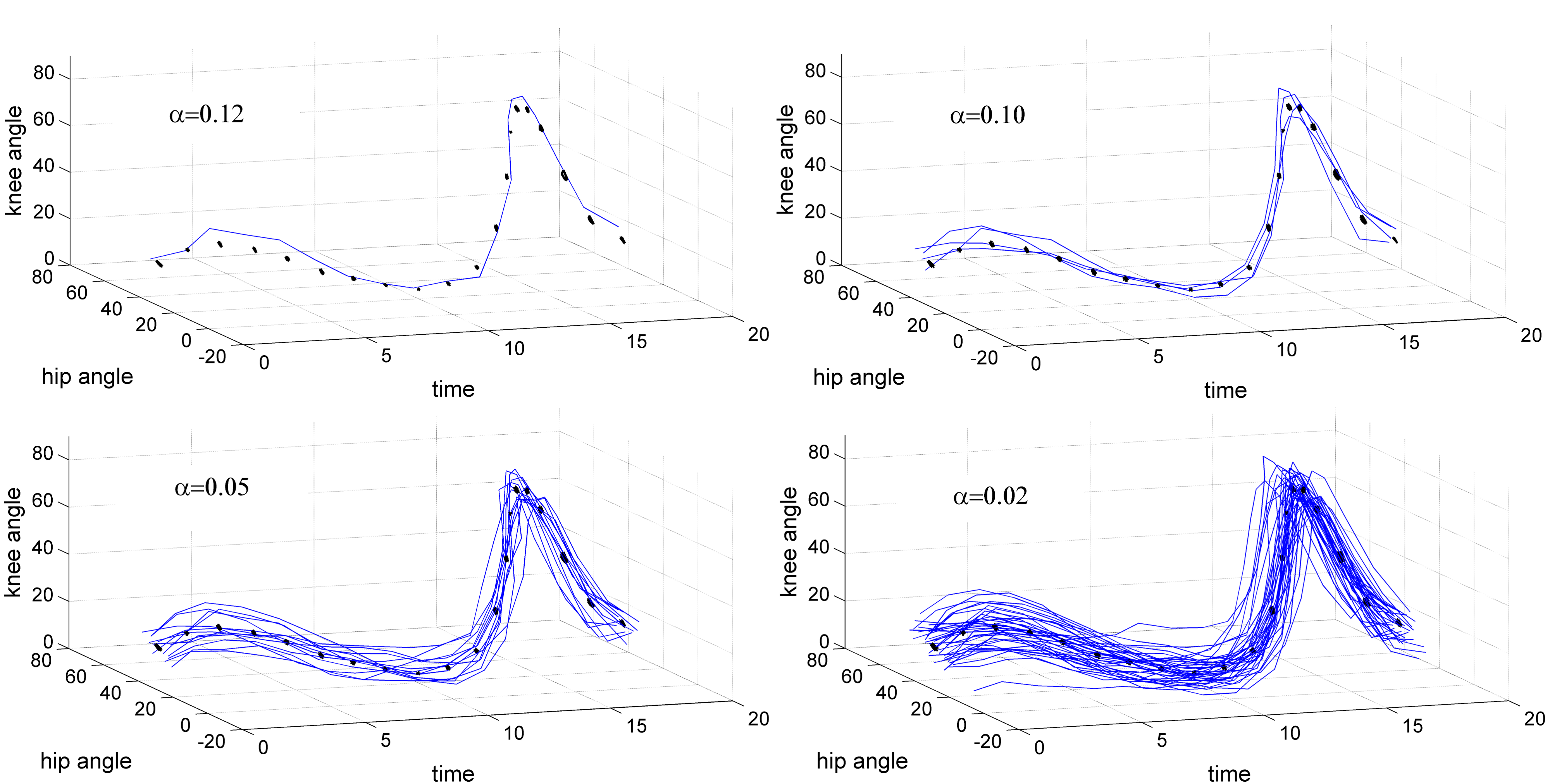}
\caption{Central regions  $D^{TG}_\alpha$, of bivariate Tukey graph depth, $\alpha\in \{0.02, 0.05, 0.10, 0.12\}$; hip and knee.
}
\end{center}
\label{fig4}
\end{figure}

Under certain conditions  the surjection property is true and hence equation (\ref{strictpropsets}), which allows us to state the following result:
\begin{proposition}
The graph depth satisfies the surjection property if the underlying multivariate depth is continuous in the data, which means that $D^d$ is continuous as a function $\mathbb{R}^d\times\mathbb{R}^{d\cdot n}\rightarrow\mathbb{R}$.
\end{proposition}

\textbf{Proof.} \quad We have to show that for all data clouds $X$, $t_*\in T$ and $y\in \IR^d$ exists some $z\in E$ with $z(t)=y$ and
\begin{equation}\label{eqsurjectiongraph}
 \inf_{t\in T} D^d(z(t) | X(t)) = D^d(z(t^*)|X(t^*))\,.
\end{equation}
However, as $z$ is a continuous function and the depth $D^d$ is continuous in the data, equation (\ref{eqsurjectiongraph}) holds for all $z$.
\hfill $\lozenge$

For example, the $d$-variate zonoid and Mahalanobis depths are continuous in the data, while the Tukey and simplicial depths are not.

\subsection{Location-slope graph depth}\label{subseclocationslope}

As the $\Phi$-depth is a multivariate functional depth, we may also apply it to a function and its derivatives. The simplest case is considering a univariate function $x$ together with its first derivative $x'$. Then the bivariate functional depth measures how similar a given function is to a cloud of functions in terms of location and slope. In the framework of general graph depths this is easily done as follows.

Consider $E=C^1(J; \IR)$ with a proper norm, e.g. $||z|| = ||z||_\infty + ||z'||_\infty$. Let

\begin{equation}
\Phi= \{\phi^t : \phi^t(x)=(x(t), x'(t))^\textsf{T}, t\in T\}
\end{equation}
for some  $T\subset J$. For $D^2$ use any bivariate depth that satisfies \textbf{D1} to \textbf{D5}. This results in the \emph{location-slope depth}
\begin{equation}
 D^{LS}(z|X)= \inf_{t\in T} D^2\left(\left(\begin{array}{c} z(t)\\ z'(t)\end{array} \right)
 	|\left(\begin{array}{c} X(t)\\ X'(t)\end{array} \right)
  \right)\,.
\end{equation}


Figure 5 
exhibits the development of the hip angle and its derivative in time separately (right panel) and as a bivariate function (left panel). For these data the location-slope graph depth has been calculated with an underlying bivariate Tukey depth.
Figure 6 
shows, for different choices of $\alpha$, the central regions of this location-slope depth.

\begin{figure}
\begin{center}
\includegraphics[scale=0.71]{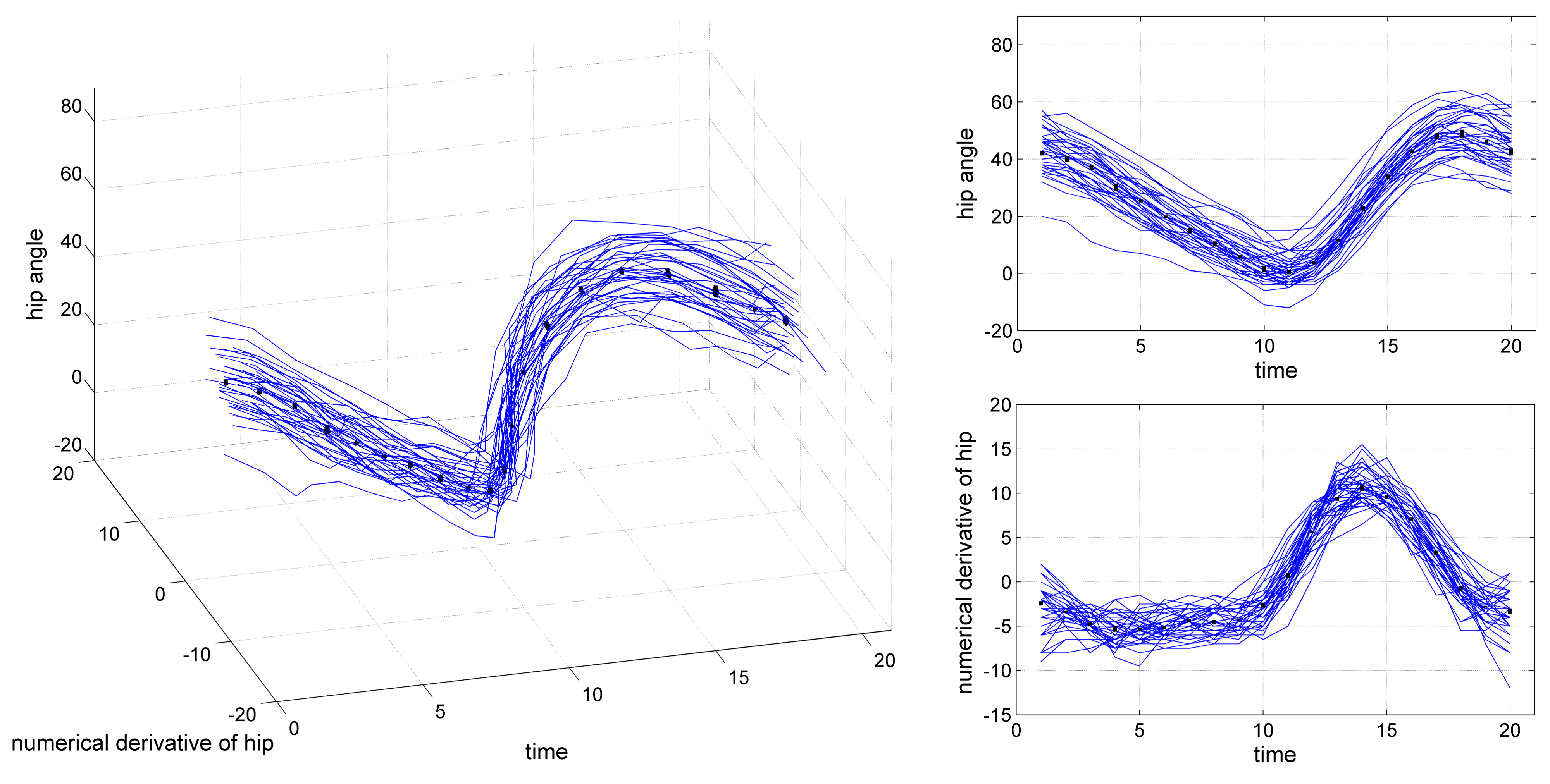}
\caption{Location and slope, univariate (right panel) and bivariate (left panel), of the hip angle.}
\end{center}
\label{fig5}
\end{figure}

\begin{figure}[h!]
\begin{center}
\includegraphics[scale=0.71]{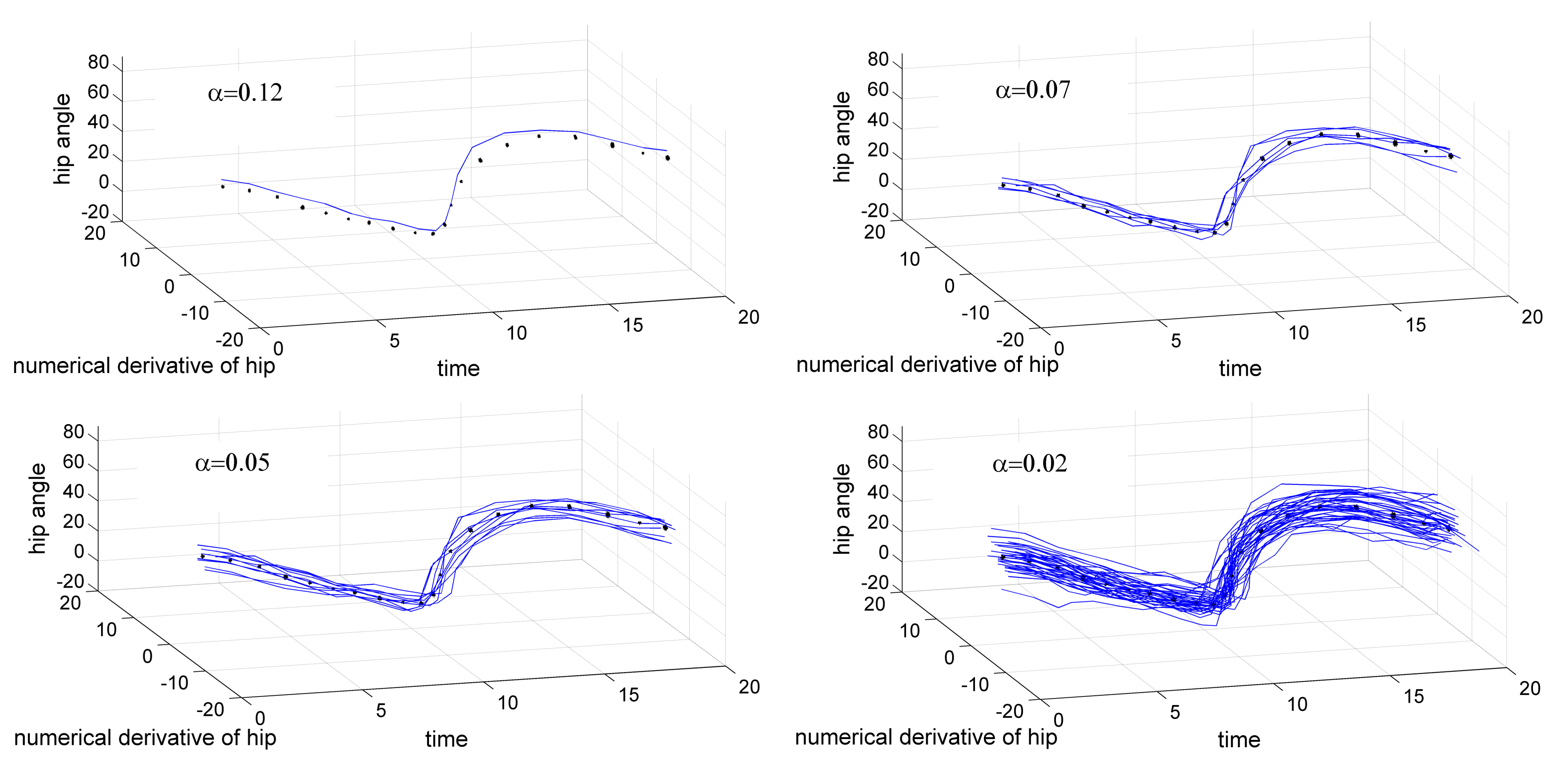}
\caption{Central regions of location-slope depth (based on bivariate Tukey depth), $\alpha\in \{0.02, 0.05, 0.10, 0.12\}$;
hip angle.}
\end{center}
\label{fig6}
\end{figure}


\begin{proposition}\label{propositionlocslope}
    The location-slope graph depth satisfies the postulates {\rm\textbf{FD1}} to {\rm \textbf{FD5}} and, in addition,  {\rm\textbf{FD2F}}. Moreover {\rm\textbf{FD2R}} and {\rm\textbf{FD2IR}} hold too if the rearrangement function $\rho$ is differentiable.
\end{proposition}

For proof see the Appendix. Here, again, $S=\{x : x(t)=0, t\not\in T\}$.
An important application of a location-slope depth is the analysis of registered functional data.
Assume that we observe time-warped functions, $x_i(t)=y_i(h_i^{-1}(t)), t\in [0,1],$ where the $y_i$ are time-synchronized and the warping functions $h_i$ are      obtained by standard procedures from the observed $x_i$; see \cite{RamsayS05}.
Then we may investigate depth and deepest points of the bivariate functions
\[\left(x_i(t),\;  h_i(t)\right), t\in [0,1]\,;\]
see also \cite{ClaeskensHSV14}.


Along the same lines we may construct $\Phi$-depths that include higher order derivatives as well as multivariate functions. In applications, by those depths it can be measured how closely a function follows the dynamics of a cloud of of functions.


\subsection{Grid depths}\label{subsecgrid}

This section introduces the grid functional depth, which is based on a special filtering of the data: The functions are evaluated on a fixed grid. While for a graph functional depth the functions are considered on the whole interval, $J$, or some subset of it, for a grid functional depth we restrict on values at some given points $t_1,\dots, t_k$ in $J$.

Let $E=C(J;\mathbb{R}^d)$ with norm $||\cdot||_\infty$. We choose a finite number of points in $J$, $t_1,\dots, t_k$, and evaluate a function $z\in E$ at these points. Notate $\underline t=(t_1,\dots, t_k)$, $z_j(\underline t)=(z_j(t_1), \dots, z_j(t_k))^\textsf{T}$, and $z(\underline t)=(z_1(\underline t), \dots, z_d(\underline t))$.  That is, in place of the $d$-variate function $z$ the $k\times d$ matrix $z(\underline t)$
 is considered. A \textbf{grid depth} $D^R$ is defined by (1) with the following $\Phi$,
\begin{equation}
\Phi= \{\phi^r : \phi^r(z)= r^\textsf{T} z(\underline t) = (r^\textsf{T} z_1(\underline t), \dots, r^\textsf{T} z_d(\underline t)), r\in S^{k-1}\}\,,
\end{equation}
which yields
\begin{equation}\label{defGrid}
 D^R(z|x^1,\dots, x^n)= \inf_{r\in S^{k-1}} D^d(r^\textsf{T} z(\underline t) | r^\textsf{T} x^1(\underline t),\dots , r^\textsf{T} x^n(\underline t))   \,.
\end{equation}
Let $S= \{ z\in E : z(t)=0 \;\; \text{for} \;\; t\not\in \{t_1,\dots,t_k\}\}$. From Theorem (1) follows:

\begin{proposition}
The class of grid depths satisfies {\rm \textbf{FD1}} to {\rm \textbf{FD5}}, with {\rm \textbf{FD3}} restricted to $||z(t^{(k)})||\to \infty$.
\end{proposition}
Obviously, a grid depth is not invariant to arbitrary or increasing rearrangements (\textbf{FD2R} or \textbf{FD2IR}), but it is invariant to permutations of
$\{t_1,\dots, t_k\}$. Also it is not function-scale invariant (\textbf{FD2F}).

When $d=1$ the grid depth can be seen as a multivariate depth $\tilde D^k$ in $\IR^k$ satisfying the weak projection property (\cite{Dyckerhoff04}),
\[
 \tilde D^k(z|x^1,\dots, x^n)= \inf_{r\in S^{k-1}} D^1(r^\textsf{T} z(\underline t) | r^\textsf{T} x^1(\underline t), \dots r^\textsf{T} x^n(\underline t))\,.
\]
In the case $d=1$ the grid depth satisfies the surjection property if and only if  for all $X, r^*$ and $y$ there is some $z\in E$ with $ r^{* \textsf{T}} z(\underline t)= y$ and
\begin{eqnarray*}
 D^R(z|X)&=&\inf_{r\in S^{k-1}} D^1(r^\textsf{T} z(\underline t) | r^\textsf{T} x^1(\underline t), \dots, r^\textsf{T} x^n(\underline t))\\
 & =& D^1(y|r^{*\textsf{T}} x^1(\underline t), \dots, r^{* \textsf{T}} x^n(\underline t)  )\,.
\end{eqnarray*}

This restriction holds, e.g., for the Mahalanobis and the zonoid depths but not for the Tukey depth; for a counterexample, see \cite{Dyckerhoff04}.


A \textbf{location-slope grid depth} is defined in the same way as the location-slope graph depth. Also higher derivatives can be included into the notion of grid depth. We omit the details.


\section{Extensions, principal component depth}\label{sec6}
More functional depths can be constructed with a  generalized versions of Definition \ref{defFunctionalDepth}.
In (\ref{defFD}) we may introduce weights $w_\phi\ge 0$ that reflect the relative importance of `direction' $\phi$, $\phi\in \Phi$. This obviously does not affect the validity of the above postulates  \textbf{FD1} to \textbf{FD5}.

\begin{definition}[Weighted functional data depth]
\begin{equation}\label{defwFD}
 D(z|X)= \inf_{\phi\in \Phi} w_\phi D^1(\phi(z)|\phi(x^1), \dots, \phi(x^n))\,.
\end{equation}
\end{definition}
Also the set $\Phi$ may be made dependent on the data. This is done in the next depth notion, the principal component depth.

Let $E$ be a separable Hilbert space, e.g.\ the space ${\cal L}^2(J)$ of all square-L-integrable functions  or the space $\ell^2$ of square-summable sequences. Given $x^1,\dots, x^n\in E$ first a -- possibly robust -- principal component (PC) analysis is performed; see \cite{RamsayS05, Shang14} and for robust methods \cite{BaliBTW11}.
Let $y_1, \dots, y_m$ denote the first $m$ eigenfunctions and
\[ q(z) = \sum_{j=1}^m \gamma_j(z) y_j
\]
be the least-squares approximation of $z\in E$. Define
\begin{equation}\label{defPhidependent}
    \Phi= \Phi(X)= \{\phi : \phi(z)=\langle r, \gamma(z) \rangle, r\in S^{m-1}\}\,,
\end{equation}
where $\gamma(z)=(\gamma_1(z),\ldots ,\gamma_m(z))$ are the scores. (In practical applications mostly $m=3$ is enough.

Obviously, given the $X$ and hence the $y_1, \dots, y_m$, the $\gamma_1,\dots,\gamma_m$ are linear and continuous in $z$. Therefore all $\phi\in \Phi$ are continuous linear functionals. We define the principal component depth as follows:

\begin{figure}[t!]
\begin{center}
\includegraphics[scale=0.57]{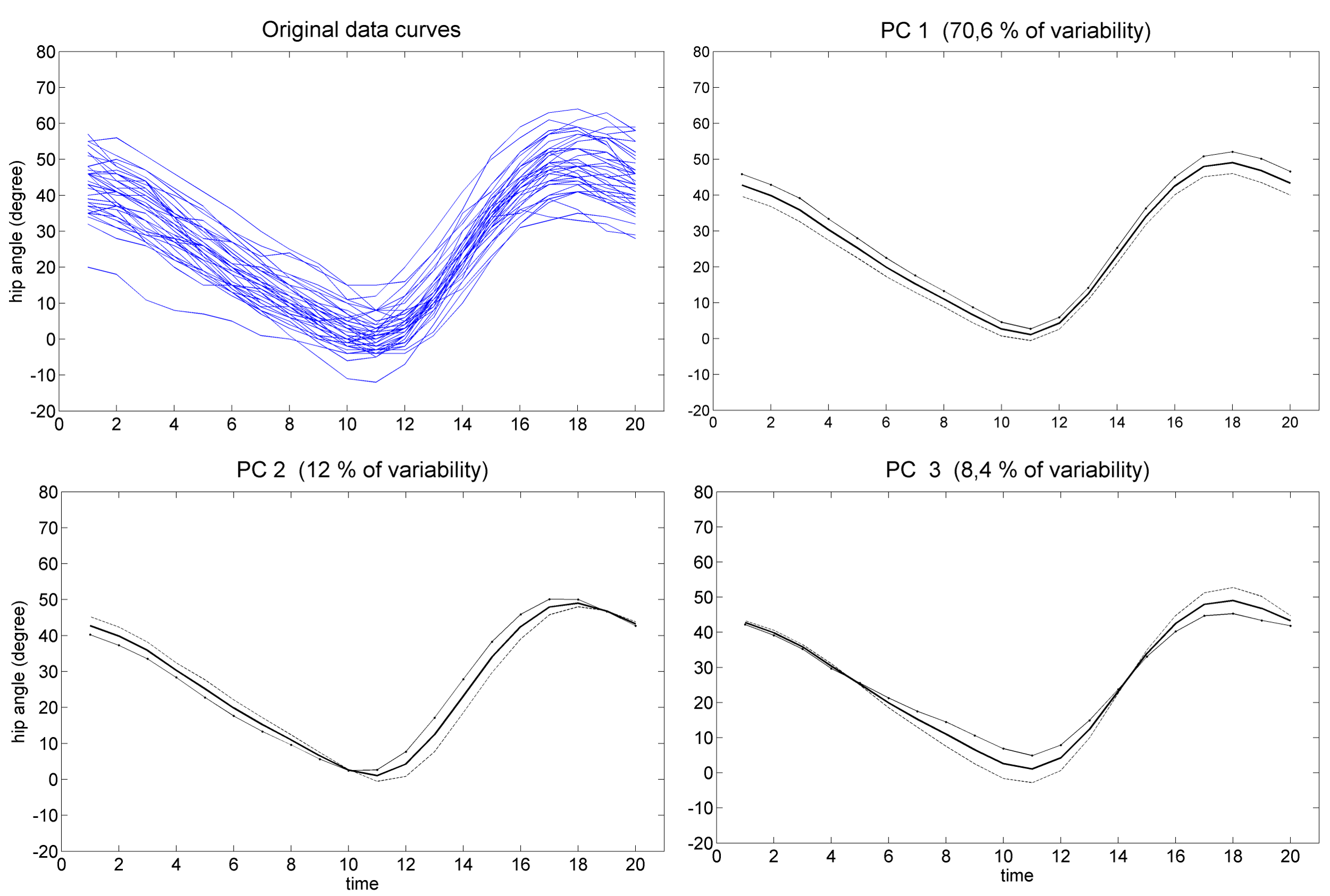}
\caption{Data and first three principal components; hip angle: Mean function (solid line), sum of (dash-dotted line) and difference between (dashed line) mean function and principal component.}
\end{center}
\label{fig7}
\end{figure}

\begin{figure}[t!]
\begin{center}
\includegraphics[scale=0.57]{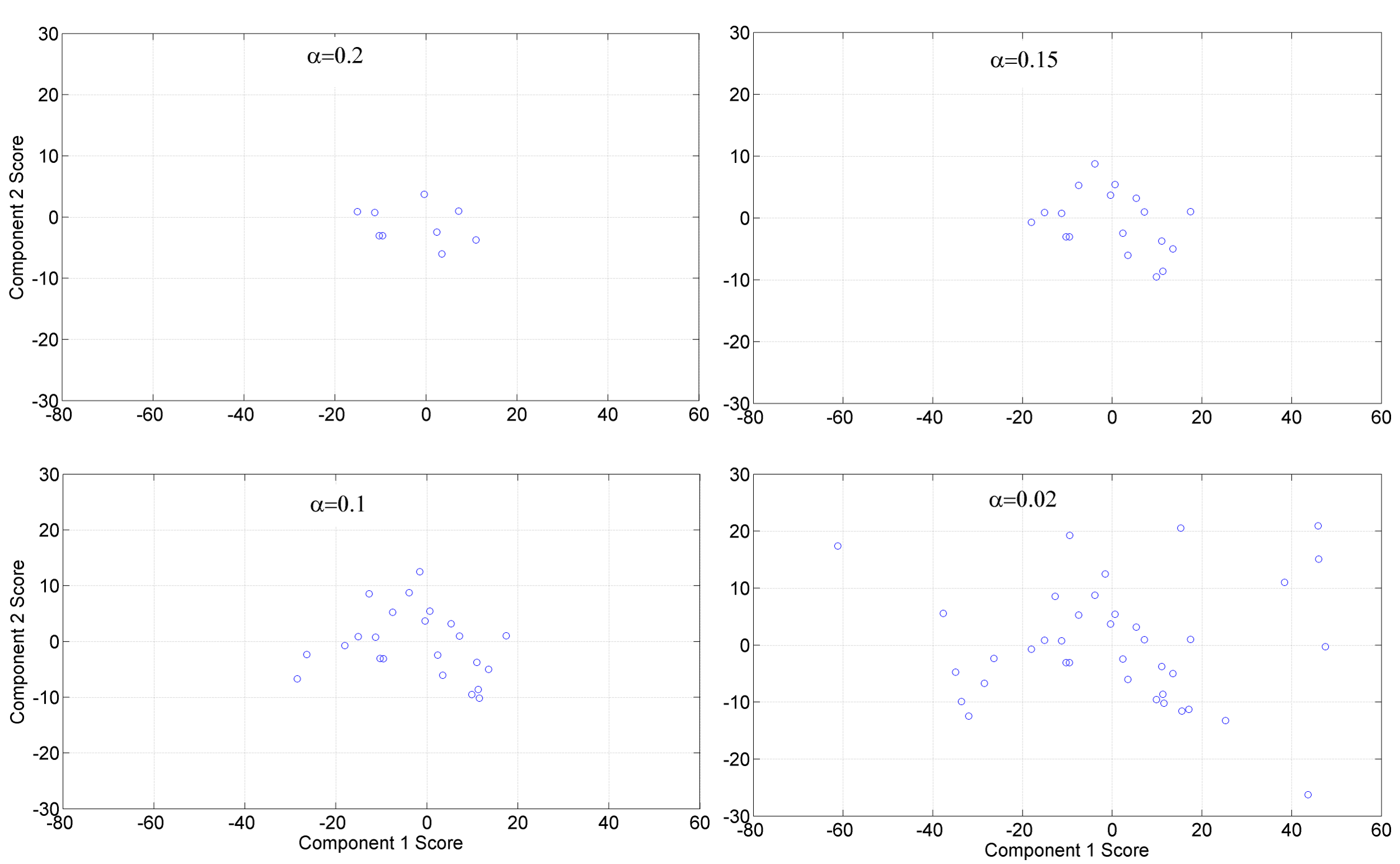}
\caption{Central regions $D^{PC}_\alpha$, $\alpha\in \{0.02, 0.10, 0.15, 0.20\}$; hip angle.
}
\end{center}
\label{fig8}
\end{figure}

\begin{definition}[Principal component depth]
\[    D^{PC}(z|X)= \inf_{r\in S^{m-1}} D^1\left(\langle r, \gamma(z) \rangle|\langle r, \gamma(x^1) \rangle, \dots, \langle r, \gamma(x^n) \rangle\right) \]
\end{definition}

Note that many multivariate depth notions, among them the location, zonoid and Mahalanobis depths, satisfy the weak projection property \citep{Dyckerhoff04}. In this case, it holds
\[  D^{PC}(z|X)=  D^m\left( \gamma(z) | \gamma(x^1), \dots, \gamma(x^n)\right)\,. \]

\begin{proposition}
 The principal component depth satisfies {\rm \textbf{FD2}},
 {\rm \textbf{FD4con}}, {\rm \textbf{FD5}}, and slight variants of {\rm \textbf{FD1}}, {\rm \textbf{FD3}} and {\rm \textbf{FD4R}}, where $b$ resp.\ $z$ resp.\ $r$ are restricted to linear combinations of the principal components, that is to elements of $\{x : x= \sum_{j=1}^m \lambda_j y_j, \lambda_j\in \mathbb{R}$.
\end{proposition}

The proof is straightforward and left to the reader.


We illustrate the PC-depth by applying it to the
hip angle data. Figure 7 
exhibits the data and its first three principal components, which are plotted as perturbations of the (pointwise) mean function. We neglect the third component and represent each function $\bmx$ by its bivariate component score $(\gamma_1(\bmx),\gamma_2(\bmx))$. Then the bivariate Tukey depth is used to construct central regions in the score space (Figure 8) 
and, consequently, as an approximation in the original data space
(Figure 9). 
Similarly, Figure 10 
shows outlying data of different maximum PC-depth. Comparing the central regions trimmed by PC-depth in  Figure 9 
with those trimmed by Tukey graph depth in Figure 2, 
we observe that at level $\alpha=0.02$ both trimmings provide the full data set, while at level $\alpha=0.02$ the PC-depth yields a larger region, which near the right border of the interval spreads significantly more out. This illustrates the different approaches: The PC-depth relates to the common principal components and measures centrality with respect to their loadings, while the Tukey graph depth refers to the functional level of the data and indicates uniform centrality over the interval.

\begin{figure}[t!]
\begin{center}
\includegraphics[scale=0.57]{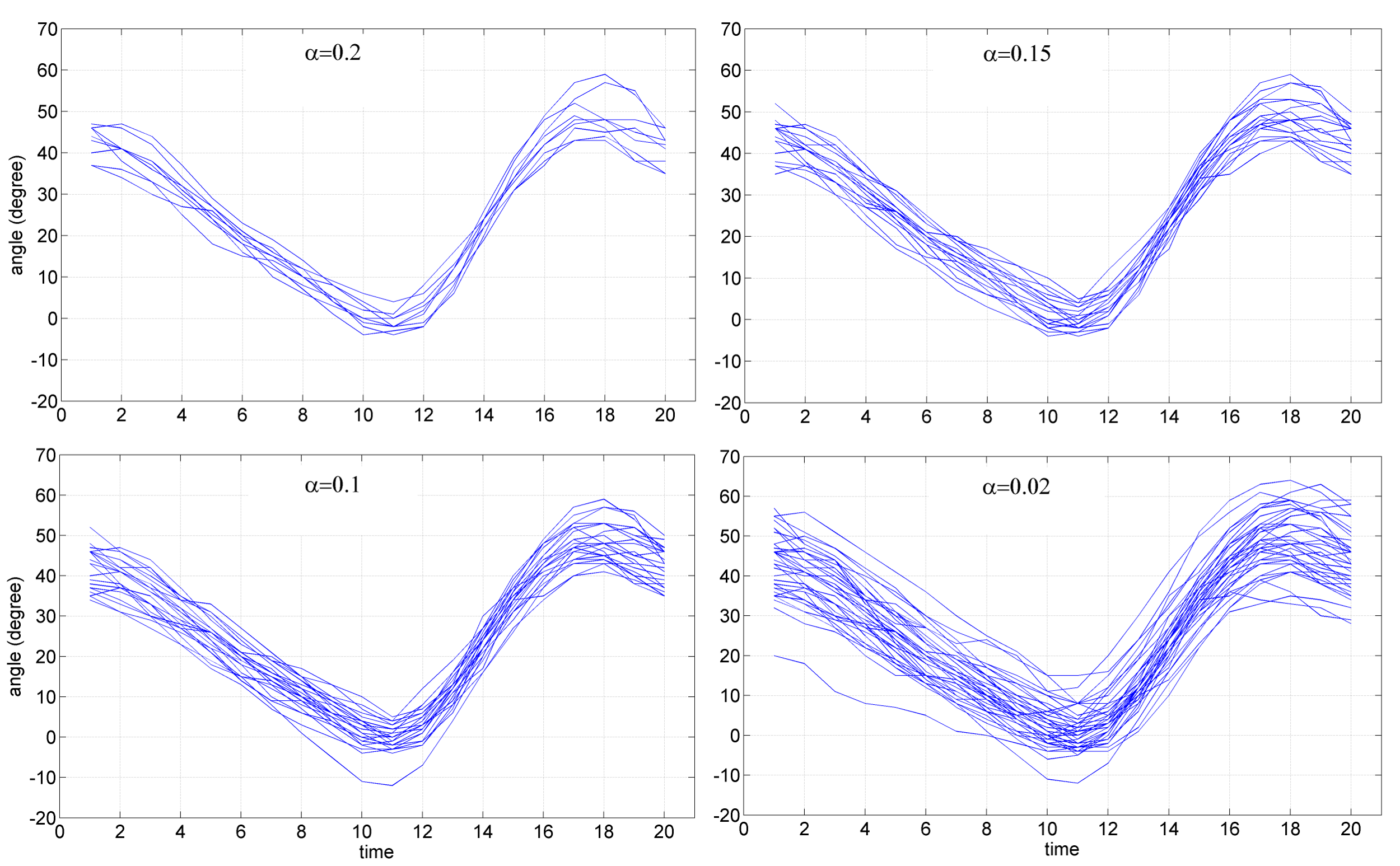}
\caption{Central regions $D^{PC}_\alpha$, $\alpha\in \{0.02, 0.10, 0.15, 0.20\}$; hip angle.
}
\end{center}
\label{fig9}
\end{figure}

\begin{figure}[t!]
\begin{center}
\includegraphics[scale=0.57]{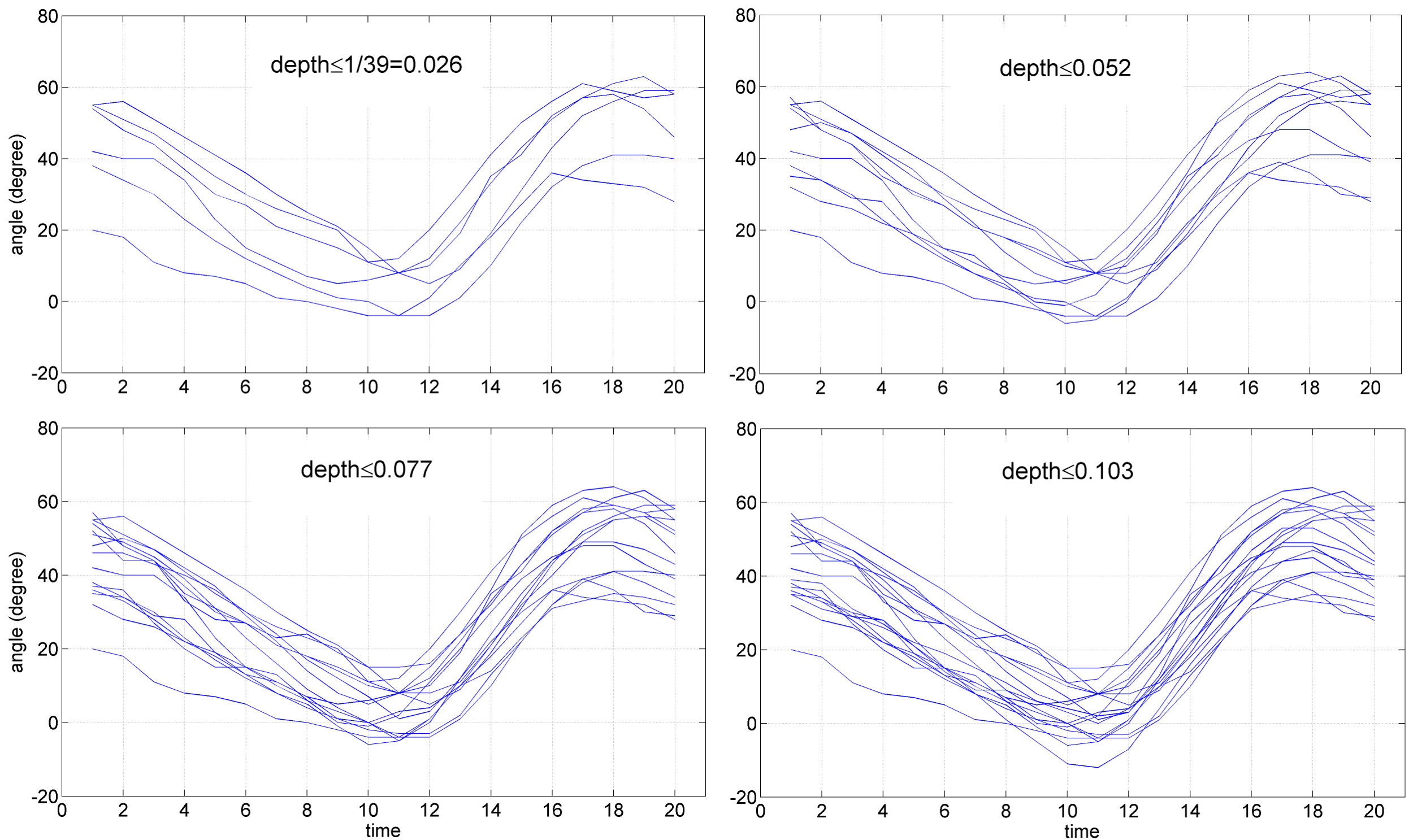}
\caption{Outlying data having $PC$-depth below
$\frac{1}{39}= 0.0256$, $\frac{2}{39}= 0.0512$, $\frac{1}{39}= 0.0769$, and $\frac{1}{39}= 0.1035$, respectively; hip angle.
}
\end{center}
\label{fig10}
\end{figure}

\section{Population versions}\label{sec6a}

Our definition \ref{defFunctionalDepth} of a $\Phi$-depth for functional data
extends immediately to a population version, that is, to a depth with respect to a probability distribution on Banach space $E$.
Note that the above postulates \textbf{FD1} to \textbf{FD5} can be literally translated to the population setting.
\begin{definition}
Let $X$ be an $E$-valued random variable, and $\Phi$ and $D^d$ as in Definition \ref{defFunctionalDepth}. The function
\begin{equation}\label{defPopulationDepth}
    D(z|X)= \inf_{\phi\in \Phi} D^d(\phi(z)|\phi(X))\,, \quad z\in E\,,
\end{equation}
is a \emph{functional depth}.
\end{definition}
It is easily seen that $D(z|X)$ is well defined for all $z\in E$ and that the functional depth satisfies the postulates \textbf{FD1} to \textbf{FD5}.
More specifically, proving \textbf{FD1}, \textbf{FD2}, \textbf{FD4}, \textbf{FD5} is straightforward. \textbf{FD3} holds as far as $||z^i||\to \infty$ implies that there exists a sequence $(\phi_i)\subset \Phi$ such that $\phi_i(z^i)\to \infty$.

However, the population versions of $\Phi$-depths are problematic, since they often collapse to zero.
\cite{ChakrabortyC14} have demonstrated that the population versions of the band depth \citep{LopezPR09}
as well as of the half-region depth become trivial, that is, almost surely equal to zero, under a broad class of standard generating probability models.
See also Section 3.3 in \cite{KuelbsZ13}, who provide an example of a $\Phi$-depth, which is trivial even with a countable $\Phi$.
Consequently, these depths must be taken as purely data-analytic tools, without reference to a generating probability distribution allowing for consistency or other asymptotics.


\section{Concluding remarks}\label{sec7}

A general framework of postulates has been given to define a depth for functional data.
We have demonstrated that the $\Phi$-depths form a comprehensive and flexible class that satisfies the basic postulates of functional data depths and contains special notions for diverse applications.

In applying the notion of $\Phi$-depth to a real data problem, we have to make two choices: selecting a proper set $\Phi$ of aspects and choosing an underlying multivariate data depth $D^d$.

The selection of the set of aspects, $\Phi$, essentially depends on the nature of the problem at hand and the goal of the analysis. There is no universally feasible choice of $\Phi$ and no all-purpose functional data depth. Specifically, to cope with a problem of functional outlier identification, we have first to discuss which features make a function an outlying one. This, e.g., can be the occurrence of local peaks, general location, either local or global growth behaviour, or a particular 'pathologic' shape \citep{Mosler15}. Similar with problems of classification: E.g., the famous and widely analyzed Berkeley data on heights of boys and girls are best classified by viewing at their growth behavior in the middle of the time interval; see \cite{MoslerM14}.
Additionally, in choosing $\Phi$, one should keep in mind that the validity of some depth postulates as well as the discriminating power of the functional depth are affected by the extensiveness of $\Phi$.

In the selection of $D^d$, questions of computability and - depending on the data situation - robustness are of primary importance. Also, as we have seen, properties like quasiconcavity (\textbf{FD4con}) and the surjection property depend on the choice of  $D^d$. Mahalanobis depth is solely based on estimates of the mean vector and the covariance matrix. In its classical form with moment estimates Mahalanobis depth is efficiently calculated but highly non-robust, while with estimates like the minimum volume ellipsoid it becomes more robust. However, since it is constant on ellipsoids around the center, Mahalanobis depth cannot reflect possible asymmetries of the data. Zonoid depth can be efficiently calculated, also in larger dimensions, but has the drawback that the deepest point is always the mean, which makes the depth non-robust. So, if robustness is an issue, the zonoid depth has to be combined with a proper preprocessing of the data to identify possible outliers. The Tukey depth is, by construction, very robust but expensive when exactly computed in dimensions $>3$. As an efficient approach the random Tukey depth can be calculated, where the minimum of univariate depths in several random directions is determined. This yields an upper bound of the Tukey depth; however the number of directions has to be somehow chosen.
Further qualified candidates, among others, are projection depths and - albeit being only
mirror symmetric - $L^p$-depths.

However, as we have pointed out in the preceding section, $\Phi$-depths often have only trivial population versions. Then, they cannot be meaningfully related to a generating probability model, and no consistency or other asymptotic results are available. Consequently, these $\Phi$-depths must be considered as purely data-analytic tools. The same holds for the half-region depth.
\cite{KuelbsZ15} have developed a method of smoothing data by perturbation, that leads to non-trivial population versions of the half-region and other depths.

Other approaches in the literature are
mainly of two types. The first type employs random projections of the data:
\cite{CuestaAN08b} define the depth of a function as the univariate depth of the function values taken at a randomly chosen argument $t$.
\cite{CuestaAN08a} propose the random Tukey depth, which is the minimum univariate Tukey depth of univariate projections in a number of random directions; with $d$-variate data the random Tukey depth converges almost surely from above to the Tukey depth.
\cite{CuevasFF07} also employ a random projection method.
The other type uses average univariate depths. Compared to this our definition may be mentioned as a `uniform' depth:
\cite{FraimanM01} calculate the univariate depths of the values of a function and integrate them over the whole interval; this results in kind of `average' depth.
\cite{ClaeskensHSV14} introduce a multivariate $d\ge 1$ functional data depth, where they similarly compute a weighted average depth. The weight at a point reflects the variability of the function values at this point (more precisely: is proportional to the volume of a depth trimmed region at the point).
These notions satisfy the above basic postulates or proper modifications of them; but a detailed analysis of them as well as a discussion of the alternative postulates recently given in \cite{NietoRB16} are beyond the scope of this paper. Recently, \cite{Nagy16} provides a comprehensive and deep investigation into notions of depth for functional data, including infimum depth; see also \cite{GijbelsN15}.



\section*{Appendix}

\textbf{Proof of Theorem \ref{theo1}.}\\
(i): The function $\phi(\cdot)=(\phi_1(\cdot),\ldots,\phi_d(\cdot))\in \Phi$ is linear, since it is linear in every component.
 \textbf{FD1} and \textbf{FD2} are obvious due to the linearity of $\phi$ and the affine invariance \textbf{D1} and \textbf{D2} of $D^d$\,.

To show \textbf{FD4}, assume that $z^*\in \bigcap_{\phi \in \Phi} \phi^{-1}(D_1^d(\phi(X)))$, in particular, that this intersection is not empty. Then
$D^d(\phi(z^*)|\phi(X))=1$ for all $\phi$, and therefore $D(z^*|X)=  \inf_{\phi\in \Phi} D^d(\phi(z^*)|\phi(X))=1$. We conclude that $z^*$ is a $D$-deepest point in $X$. It holds
	\begin{eqnarray*}
  D(z^*+\alpha r|X) &=& \inf_{\phi\in \Phi} D^d(\phi(z^*+\alpha r)|\phi(X))\\
  					&=& \inf_{\phi\in \Phi} D^d(\phi(z^*)+\alpha \phi(r)|\phi(X)).
	\end{eqnarray*}
	Since $D^d(\phi(z^*)+\alpha \phi(r))|\phi(x^1), \dots, \phi(x^n))$ decreases with $\alpha>0$ according to \textbf{D4}, \textbf{FD4} is true.
	
For \textbf{FD5}, note that
\begin{eqnarray*}
   D_\alpha(x^1,\dots, x^n)&=& \{z : D(z|X)\ge \alpha\}=  \{z : \inf_{\phi\in\Phi} D^d(\phi(z)|\phi(X))\ge \alpha\}\\
   &=&  \bigcap_{\phi\in\Phi} D^d_\alpha (\phi(X))\,,
\end{eqnarray*}
which, as an intersection of closed sets, is closed.

(ii): Obvious.

(iii):
To show \textbf{FD4con}, assume $y, z\in D_\alpha(X)$, hence, for all $\phi\in \Phi$,
\[D^d(\phi(y)|\phi(X)) \ge \alpha \quad \text{and} \quad  D^d(\phi(z)|\phi(X)) \ge \alpha\,.\]
For every $\lambda\in ]0.1[$ follows (due to the linearity of $\phi$ and \textbf{D4con}):
\begin{equation*}
D^d(\phi(\lambda y + (1-\lambda) z)|\phi(X)) =  D^d(\lambda \phi(y) + (1-\lambda) \phi(z)|\phi(X))\ge \alpha\,.
\end{equation*}
By taking the infimum, conclude that $D(\lambda y + (1- \lambda) z|X)\ge \alpha$; therefore \textbf{FD4con
}.
\hfill $\lozenge$

\textbf{Proof of Theorem \ref{weaksets}.} \quad First assume that $z\in D_\alpha(X)$, i.e.,
\[D(z|X) =\inf_{\phi\in \Phi} D^d(\phi(z)|\phi(X))\ge \alpha\,.
\]
Then, for every $\phi\in\Phi$, obtain $D^d(\phi(z)|\phi(X))\ge \alpha$, hence $\phi(z)\in D^d_\alpha(\phi(X))$, and therefore $z\in \phi^{-1}(D^d_\alpha(\phi(X)))$.  Conclude
\[D_\alpha(X) \subset \bigcap_{\phi\in\Phi} \phi^{-1}(D^d_\alpha(\phi(X)))\,.
\]
On the other hand, let $z\in \bigcap_{\phi\in\Phi} \phi^{-1}(D^d_\alpha(\phi(X)))$, which means that for all $\phi\in \Phi$ exists some $y\in D^d_\alpha(\phi(X))$ with $y=\phi(z)$ and $D^d(y|\phi(X))\ge \alpha$. Consequently, $D(z|X) \ge \alpha$, and $z\in D_\alpha(X)$, which proves
the reverse set inclusion, hence Equation (\ref{weakpropsets}).
\hfill $\lozenge$

\textbf{Proof of Theorem \ref{theosurjection}.} \quad `only if': Let $\phi\in \Phi$. From Equation (\ref{weakpropsets}) follows that $D_\alpha(X) \subset \phi^{-1}(D^d_\alpha(\phi(X)))$, and therefore
\begin{equation}\label{strictpropsubset}
\phi(D_\alpha(X)) \subset D^d_\alpha(\phi(X))\,.
\end{equation}
Now let $y\in D^d_\alpha(\phi(X))$. Then, by the surjection property exists some $z$ with $\phi(z)=y$ and
\[ D(z|X)=D^d(y|\phi(X))\ge \alpha \,.
\]
Hence $z\in D_\alpha(X)$, and therefore $y=\phi(z)\in \phi(D_\alpha(X))$. We conclude $ D^d_\alpha(\phi(X))\subset \phi(D_\alpha(X))$ and finally, with (\ref{strictpropsubset}) the claimed equality (\ref{strictpropsets}).\\
`if': Assume that (\ref{strictpropsets}) holds and let $y\in \IR^d$, $\alpha=D^d(y|\phi(X))$. Then, for all $\phi\in \Phi$
\[ y\in D^d_\alpha(\phi(X))= \phi(D_\alpha(X))
\]
and there exists $z\in D_\alpha(X)$ so that $y=\phi(z)$. By this and the defining Equation (\ref{defFunctionalDepth}) it follows that
\[ \alpha\le D(z|X) \le D^d(\phi(z)|\phi(X))=\alpha\,,
\]
and therefore $D(z|X) \le D^d(\phi(z)|\phi(X))=\alpha$, which proves the surjection property.
\hfill $\lozenge$

\textbf{Proof of Proposition \ref{propositiongraph}.} For \textbf{FD1}, \textbf{FD2}, \textbf{FD4} and \textbf{FD5} see Theorem \ref{theo1}.
Let $S=(\mathbb{R}^d)^T$, which is the subspace of functions $T\to \mathbb{R}^d$.
Consider $z^i \in S$, $i\in \mathbb{N}$, with $\lim||z^i||=\lim_i \sup_{t\in T} z^i(t)=\infty$. Then there exists a sequence $t_i\in T$, $i\in \mathbb{N}$, so that $\lim_i z^i(t_i)= \lim_i \phi^{t_i}(z^i)= \infty$. Hence \textbf{FD3} holds with this subspace $S$.

To show \textbf{FD2F}, let $a\in E_0$ and consider the componentwise multiplication of $a$ with some $z\in E$,
$$(a\cdot z)(t)=a(t)\cdot z(t)=(a_1(t) z_1(t), \ldots , a_d(t) z_d(t))^{T}= A(t)(z(t))\,, $$
where $A(t)=diag (a_1(t), a_2(t), \ldots , a_d(t))$ is a regular matrix. Notate similarly $(a\cdot X)(t)=A(t) X(t)$.
Then, by \textbf{D2}, it holds $D^d(A(t) z(t) | A(t) X(t))=D^d(z(t) | X(t))$ for every $t\in T$ and hence
\begin{eqnarray*}
D^G(a\cdot z|a\cdot X)&=& \inf_{t\in T} D^d((A(t) z(t)| A(t) X(t))\\
	&=&\inf_{t\in T} D^d(z(t)| X(t))\\
	&=& D^G(z|X)\,.
\end{eqnarray*}
Next we demonstrate \textbf{FD2R}. Let $\rho$ be a bijection on $J$, $s=\rho(t)$.
\begin{eqnarray*}
D^G(z\circ \rho|x^1\circ \rho,\dots, x^n\circ \rho)&=& \inf_{t\in T} D^d( z(\rho(t))| x^1(\rho(t)),\dots,  x^n(\rho(t)))\\
&=& \inf_{s\in T} D^d( z(s)| x^1(s),\dots,  x^n(s))\\
&=& D^G(z|x^1,\dots, x^n)\,.
\end{eqnarray*}
Obviously, \textbf{FD2IR} follows.
\hfill $\lozenge$

\textbf{Proof of Proposition \ref{propositionlocslope}.} For \textbf{FD1} to \textbf{FD5} see Theorem \ref{theo1}.
Regarding \textbf{FD2F} consider
$$A(t)=\left(
\begin{array}[]{cc}
a(t)  &	0\\
a'(t)	& a(t) 		
\end{array} \right)$$
for some $a\in E$. Then
$$A(t)\left(\begin{array}{c} x(t)\\ x'(t) \end{array} \right) =
\left(\begin{array}{c} a(t) x(t)\\a'(t) x(t)+ a(t) x'(t) \end{array} \right)$$
holds for every $x\in E$.
This implies
\begin{eqnarray*}
&&D^{LS}(a\cdot z|a\cdot x^1,\dots, a\cdot x^n)\\
&=& \inf_{t\in T} D^2\left(\left(\begin{array}{c} (a\cdot z)(t)\\ (a\cdot z)'(t)\end{array} \right)
 	|\left(\begin{array}{c} (a\cdot x^1)(t)\\ (a\cdot x^1)'(t)\end{array} \right),\dots,
 		\left(\begin{array}{c} (a\cdot x^n)(t)\\ (a\cdot x^n)'(t)\end{array} \right)\right)\\
&=& \inf_{t\in T} D^2\left(A(t)\left(\begin{array}{c} z(t)\\ z'(t) \end{array} \right)|
(A(t)\left(\begin{array}{c} x^1(t)\\ {x^1}'(t) \end{array} \right),\dots,
A(t)\left(\begin{array}{c} x^n(t)\\ {x^n}'(t) \end{array} \right)\right)\\
&=& \inf_{t\in T} D^2\left(\left(\begin{array}{c} z(t)\\ z'(t) \end{array} \right)|
\left(\begin{array}{c} x^1(t)\\ {x^1}'(t) \end{array} \right),\dots,
\left(\begin{array}{c} x^n(t)\\ {x^n}'(t) \end{array} \right)\right)\\
&=&D^{LS}( z| x^1,\dots,  x^n),
\end{eqnarray*}	
since the property \textbf{D2} holds for the bivariate depth $D^2$.
To show \textbf{FD2R} (and \textit{a fortiori} \textbf{FD2IR}) define
$P(t)=\left(
\begin{array}[]{cc}
1 &	0\\
0	& \rho '(t) 		
\end{array} \right)$.
By assumption it holds $\rho '(t)\not=0$.
Then, due to \textbf{D2} and  $\rho(\cdot)$ being bijective, $s=\rho(t)$,
\begin{eqnarray*}	
&&D^{LS}(z\circ\rho| x^1\circ\rho,\dots,x^n\circ\rho)\\
&=& \inf_{t\in T} D^2\left(\left(\begin{array}{c}  z(\rho(t))\\  (z(\rho(t)))'\end{array} \right)
 	|\left(\begin{array}{c} x^1(\rho(t))\\ (x^1(\rho(t)))'\end{array} \right),\dots,
 		\left(\begin{array}{c} x^n(\rho(t))\\ (x^n(\rho(t)))'\end{array} \right)\right)\\
&=& \inf_{t\in T} D^2\left(P(t)\left(\begin{array}{c} z(\rho(t))\\ z'(\rho(t)) \end{array} \right)|
P(t)\left(\begin{array}{c} x^1(\rho(t))\\ {x^1}'(\rho(t)) \end{array} \right),\dots,
P(t)\left(\begin{array}{c} x^n(\rho(t))\\ {x^n}'(\rho(t)) \end{array} \right)\right)\\
&=& \inf_{t\in T} D^2\left(\left(\begin{array}{c} z(\rho(t))\\ z'(\rho(t)) \end{array} \right)|
\left(\begin{array}{c} x^1(\rho(t))\\ {x^1}'(\rho(t)) \end{array} \right),\dots,
\left(\begin{array}{c} x^n(\rho(t))\\ {x^n}'(\rho(t)) \end{array} \right)\right)\\
&=& \inf_{s\in T} D^2\left(\left(\begin{array}{c} z(s)\\ z'(s) \end{array} \right)|
\left(\begin{array}{c} x^1(s)\\ x'_1(s) \end{array} \right),\dots,
\left(\begin{array}{c} x^n(s)\\ x'_n(s) \end{array} \right)\right)\\
&=&D^{LS}( z| x^1,\dots,  x^n).
\end{eqnarray*}	
This completes the proof.
\hfill $\lozenge$

\subsubsection*{Acknowledgement} Thanks are to Rainer Dyckerhoff, Dominik Liebl, Mia Hubert, Gerda Claeskens, Pauliina Ilmonen, Pavlo Mozharovskyi, Pavel Bazovkin, and Stanislav Nagy for many discussions and useful comments on previous versions of the paper.

\end{document}